\documentclass[useAMS]{article}
\usepackage[intlimits]{amsmath}
\usepackage{amssymb}
\usepackage[cp1251]{inputenc}
\usepackage[T2A]{fontenc}
\usepackage[russian,english]{babel}
\usepackage{graphicx}
\usepackage{amsfonts}

\begin{document}
\voffset=-0.5 in

\title{\textbf{On the possible observational manifestation of supernova shock impact on the neutron star magnetosphere}}

\author{A.E. Egorov, K.A.Postnov
\thanks{E-mail: astrotula@gmail.com(AEE);
kpostnov@gmail.com(KAP)}\\
\small Sternberg Astronomical Institute, Universitetski pr. 13, \small  Moscow
119992, Russia}

\date{
}


\maketitle

\begin{abstract}
Impact of supernova explosion on the neutron star magnetosphere in
a massive binary system is considered. The supernova shock
striking the NS magnetosphere filled with plasma can lead to the
formation of a magnetospheric tail with significant magnetic
energy. The magnetic field reconnection in the current sheet
formed can convert the magnetic energy stored in the tail into
kinetic energy of accelerated charged particles. Plasma
instabilities excited by beams of relativistic particles can lead
to the formation of a short pulse of coherent radio emission with
parameters similar to those of the observed bright extragalactic
millisecond radio burst (Lorimer et al. 2007).
\end{abstract}

Subject headings: stars: neutron --- radio: bursts --- pulsars: general


\section{Introduction}
\label{sec:intro}

Supernova (SN) explosion in a binary system with neutron star (NS) is not a rare event. As the binary system can be disrupted during
the second SN explosion, the lower limit to the rate of such events
can be estimated, for example, from the galactic binary NS formation rate. This rate as inferred from binary pulsar statistics and population synthesis
calculations is of order of $10^{-4}-10^{-5}$ per year (Postnov \& Yungelson 2006),
i.e. a few -- ten events per cubic Gpc per day.

The impact of the SN shock on NS magnetosphere can lead to
spectacular observational manifestations. For example,
Istomin \& Komberg (2002) argue that it can result in a narrow-collimated jet
of hard radiation which can be observed as a gamma-ray burst
(GRB). Their arguments are based on the cumulative effect of the
SN shock acting on the NS magnetosphere resulting in the formation of a long
magnetospheric tail. The magnetic field reconnection in the tail
then gives rise to particle acceleration. Synchrotron emission of
these particles in the magnetic field can produce hard X-ray
photons observed as a gamma-ray burst. Conversion of magnetic
energy in the tail into hard X-rays is a complicated physical
problem. It is still to be seen if this mechanism can explain the
gamma-ray burst phenomenon. However the generation of relativistic
particles during magnetic field reconnection seems to be
unavoidable (see recent numerical calculations in
(Zenitani \& Hoshino 2007)), and a large variety of plasma phenomena are
expected to occur during interaction of these particles with the
ambient medium.

Here we suggest that plasma instabilities in the process of the
magnetic field reconnection can be responsible for the formation
of a powerful short non-thermal radio burst. A very short radio
burst with peculiar properties was recently discovered by
Lorimer et al. (2007). The radio flux of the burst at frequency 1.4 GHz
was $30\pm 10$ Jy, the duration of the event was estimated to be
shorter than 5~msec and its dispersion measure was 375 cm$^{-3}$pc
suggesting the possible extragalactic nature of the source at a distance of
$\sim$~1~Gpc. No host galaxy up to the 18th $B$-magnitude has been
found implying a distance limit of $>$600~Mpc for a Milky Way-like
host. Such a distance implies a radio energy release  in the burst
of $\sim 10^{40}$ erg with a brightness temperature of $\sim
10^{34}$~K for isotropic radiation. The total rate of such
bursts was estimated roughly to be around 90~d$^{-1}$~Gpc$^{-3}$, which is
much lower than the core-collapse SN rate
$\sim$~1000~d$^{-1}$~Gpc$^{-3}$, but well in excess of the GRB
rate $\sim$~4~d$^{-1}$~Gpc$^{-3}$ and the expected rate of binary
neutron star coalescences $\sim$~2~d$^{-1}$~Gpc$^{-3}$. This 
estimate, of course, is very uncertain due to a small statistics.

We should
note here that properties of the short radio burst reported in
Lorimer et al. (2007) are in fact very similar to those of 
pulsar radio emission:
the coherence, an extremely high brightness temperature at frequency
$\nu\approx1.4$ GHz, a steep spectral shape ($\sim \nu^{-4}$).
Lorimer et al. (2007) discuss several possible known sources of the
strong millisecond radio bursts, including rotating radio
transients (RRATs) and giant pulses from radio pulsars, but none
of them appear to be energetic enough. Based on statistical
arguments only, Popov \& Postnov (2007) proposed that this burst could be
produced by a hyperflare of an extragalactic magnetar. Also it was estimated in (Popov \& Postnov 2007) that assuming the ratio of radio emission power to the
magneto-rotational losses as observed in giant radio pulses in
Crab with a peak flux of 1000 Jy, a simple scaling leads to the
conclusion that magnetar magnetic fields ($B \sim 10^{15}$~G) and
millisecond rotation periods are required. However, magnetars are observed
to rotate slowly $P=1-10$~s due to rapid spin-down. In the ms
magnetar model (Usov 1992, Usov 1994) the characteristic
spin-down time is several seconds, so if radio emission were
produced in a newly born magnetar by some mechanism, we would
observe a series of ms pulses with decreasing period, not a single bright
burst.

In contrast, the magnetic field reconnection during relaxation of NS magnetosphere after impact of SN shock naturally leads to the formation of relativistic particle streams. Plasma instabilities in such flows stimulate conversion of kinetic energy of particles to the energy of transverse electromagnetic waves, which can be observed as the short powerfull burst of non-thermal radio emission. Lyubarsky (2008b)
has shown that such radio bursts can propagate through the stellar
wind without being destroyed by the induced Compton and Raman scattering provided that they originate at distances $>10^{15}$~cm from the
exploding massive star and their source is moving relativistically. While in
massive binary systems such distances from
the SN progenitor may be natural, the relativistic bulk motion of the emitting region requires justification. In our model emitting region moves relativistically due to acceleration of electrons during non-stationary magnetic reconnection.

\section{Shock interaction with NS magnetosphere}

Following Istomin \& Komberg (2002), we consider typical SN shock parameters
at the free expansion stage after the explosion (Imshennik \& Nadyozhin 1989):
the total kinetic energy $E_{kin}  \approx 10^{47}$ erg, the
velocity $u \approx 4\cdot10^9$~cm/s, the density
$\rho \approx  10^{-8}$ g/cm$^3$, and the shock width
$h \approx 10^9$~cm.

Assume that the NS was a pulsar just before the explosion, i.e. relativistic plasma was generated near its surface. A 
young NS can shine as radio pulsar for about $10^7$~yrs. In a massive
binary, the probability for a SN explosion during this stage is
not small, especially if the original binary components had close
masses. The characteristic size of the NS magnetosphere (ignoring
possible asymmetry due to a strong stellar wind of the
secondary component) is about the light cylinder size $R_{l}\sim
c/\Omega=cP/2\pi$. The NS magnetosphere must be filled with pair plasma
with the number density determined by the Goldreich-Julian value
(Goldreich \& Julian 1969):
\begin{equation}
n_{GJ}=\frac{\left|\rho_{GJ}\right|k}{e}=\frac{|\mathbf{\Omega}\mathbf{B}|k}{2\pi ce}\,,
\label{GJ}
\end{equation}
where $k=10^4-10^5$ is the particle multiplication factor, $e$ is the  elementary charge.

The shock kinetic energy geometrically intercepted by the NS magnetosphere can be estimated as
\begin{equation}
E' = E_{kin}\frac{\pi R_{l}^2}{4\pi a^2} = 10^{39}\left(\frac{\Omega}{10\mbox{rad/s}}\right)^{-2}\left(\frac{a}{\mbox{1AU}}\right)^{-2}\mbox{erg},
\end{equation}
where $a$ is the distance between the stars. According to
Istomin \& Komberg (2002), such shock perturbation can lead to the formation of a
magnetospheric tail with substantial magnetic energy.
The tail is unstable: magnetic reconnection occurs
and charged particles are accelerated to high energies. We suggest
 that the disruption of the magnetospheric tail can lead to
the formation of two colliding relativistic electronic and
positronic beams. Plasma instabilities developed
in the collision generate
plasma waves which further can be converted into transverse
electromagnetic waves, giving rise to a narrow-collimated beam of
outgoing non-thermal radio emission. The situation here is similar to the
most plausible mechanism of radio emission generation in pulsars
(Lyubarsky 1992, 2008a).

Istomin \& Komberg (2002) argue that SN shock passing through the NS
magnetosphere compresses its head-on part and forms a long narrow
magnetospheric tail like in the terrestrial magnetosphere blown by
solar wind. In such a magnetospheric tail
conditions of the
magnetic field freezing into the shock plasma and the magnetic flux conservation in the tail (see Istomin \& Komberg (2002) for more detail)
imply
that the magnetic field strength in the tail and the tail's diameter
at distance $l$ from the NS are
\begin{equation}
B_{t}=B^{*}\left(\frac{l}{r^{*}}\right)^{1/2}\,,
\end{equation}
\begin{equation}
d_{t}=2r^{*}\left(\frac{l}{r^{*}}\right)^{-1/4}\,,
\end{equation}
respectively. Here 
	\[
B^{*}=(4\pi \rho u^2)^{1/2} \approx 10^6\left(\frac{\rho}{10^{-8}\mbox{g/cm}^3}\right)^{1/2}\left(\frac{u}{4\cdot10^9\mbox{cm/s}}\right)\mbox{G}
\]
is the magnetic
field at the Alfvenic radius $r^*$ determined as a distance where
the magnetic field pressure is balanced by the shock ram pressure.
For the dipole magnetic field and the assumed shock parameters we
find
\begin{equation}
r^*=R\left(\frac{B}{B^*}\right)^{1/3}=R\left(\frac{B^2}{4\pi \rho u^2}\right)^{1/6}\approx10^8\left(\frac{R}{10 \mbox{km}}\right)\left(\frac{B}{10^{12} \mbox{G}}\right)^{1/3}\mbox{cm}
\end{equation}
where $B$ is the field strength at the NS surface,
$R$ is the NS radius.
So the magnetic energy storage in the narrow ($d_{t} \ll L\simeq h$, where $L$ is the tail
length), magnetized ($B^*\approx10^6$ G) magnetospheric tail
will be
\begin{multline}
\epsilon_{B}=\int\frac{B_{t}^2}{8\pi}\frac{\pi d_{t}^2}{4}dl=\frac{1}{12}B^{*2}r^{*3/2}h^{3/2}=3\cdot10^{36}\left(\frac{B^*}{10^6\mbox{G}}\right)^2\left(\frac{r^*}{10^8\mbox{cm}}\right)^{3/2}\left(\frac{h}{10^9\mbox{cm}}\right)^{3/2}\mbox{erg}.
\end{multline}
Clearly, such a tail can be unstable with respect to magnetic
reconnection in the current sheet. The symmetry of the problem
suggests the plane geometry of the sheet in the orbital plane.
During the magnetic reconnection, the electric field component
transverse to the tail axis is generated in the plane sheet. This
field can accelerate plasma electrons with a broad energy
distribution up to high Lorentz factors (see e.g.
Zenitani \& Hoshino (2007)).
Importantly, some amount of electron-positron plasma is initially
expected to be present in the tail with the number density below
the Goldreich-Julian value (\ref{GJ}), since during the formation
of the magnetospheric tail electron-positron plasma will outflow
along the magnetic lines that have been opened up by the shock.
The upper limit on the number density of pair plasma in the tail
is
\begin{equation}
n_{pairs}\leq n_{GJ}(l)\approx10^7\left(\frac{\Omega}{10\mbox{rad/s}}\right)\left(\frac{B}{10^{12}\mbox{G}}\right)\left(\frac{k}{10^4}\right)\left(\frac{l}{5r^*}\right)^{-3} \mbox{cm$^{-3}$}.
\label{npairs}
\end{equation}
At the same time, some amount of electron-ion plasma can penetrate
from the shock into the tail during its formation. The number
density of electron-ion plasma can significantly exceeds that of
pairs.

\section{Generation of radio emission}

At the beginning of the magnetic reconnection, the tail is filled
with plasma with admixture of pairs. During the reconnection,
several X-points can be formed in the long tail (tearing
instability) and a strong electric field directed transversely to
the tail axis appears and accelerate charges.
Electrons and positrons are accelerated in opposite directions so
a low-density positron beam can appear propagating against
relativistic electron plasma. The situation becomes similar to the
pulsar wind flowing along the open field lines of the NS
magnetosphere (Lyubarsky 1992), where tge electric field appears due to leak of Goldreich-Julian charge density (quasi-stationary violation of condition of full screening of electric field).
A fraction of the kinetic energy of accelerated particles
can be
transformed by plasma instabilities
into plasma oscillations mostly at
the comoving plasma frequency. These
oscillations can then be non-linearly converted into transverse
electromagnetic waves
observed as pulsar radio emission.

In our case, the duration of the radio pulse produced during magnetic reconnection (fast reconnection with Alfven velocity) can be roughly
estimated from the transverse size of the emitting tail:
$\tau \approx r^*/v_{a} \approx r^*/c \approx 3$ ms.
This time is about as
observed in the reported extragalactic radio burst
(Lorimer et al. 2007).

The subsequent picture can be as follows. During the magnetic
field topology changing, a stream of relativistic electronic
plasma moving transversely to the tail axis with some effective mean Lorentz-factor $\Gamma$ can
emerge. The reverse flux of positronic plasma accelerated by the
same electric field in the reconnecting layer moves with about the
same Lorentz-factor against the electrons. Due to various instabilities (beam instability etc.) the kinetic energy of the
positronic beam dissipates in the electronic plasma by exciting
plasma oscillations. Such energy conversion is effective in the case of small number density of positrons in comparison with electrons ($n^{+}\ll n^{-}$), which is expected in our case. These oscillations are non-linearly converted
into transversal electromagnetic waves which can be observed as a
short radio burst for an appropriate orientation of the tail and the
observer. Relativistic aberration confines radio emission within a
narrow cone with opening angle $\theta \sim \Gamma^{-1}$. The
oscillations occur at the comoving plasma frequency
\begin{equation}
\tilde{\omega}=\sqrt{\frac{4\pi e^2\tilde{n^-}}{m}}.
\end{equation}
The conversion rate of plasma oscillations into transversal waves
can be expressed as (e.g. Lyubarsky (1992))
\begin{equation}
\frac{\partial \tilde{W_{t}}}{\partial t} = \tilde{\alpha}\tilde{W_{t}}\tilde{W_{l}},
\end{equation}
where
\begin{equation}
\tilde{\alpha} = \frac{\tilde{\omega}}{\tilde{n^{-}}mc^2},
\end{equation}
$\tilde{W_{t}}$ and $\tilde{W_{l}}$ are the energy
densities of the transversal and longitudinal waves,
respectively, in the comoving plasma frame.
The observed frequency is Doppler boosted (emitting electrons move towards the observer with Lorentz-factor $\Gamma$):
\begin{equation}
\nu \cong 2\Gamma \tilde{\nu}=2\Gamma \frac{\tilde{\omega}}{2\pi}=2\Gamma e\sqrt{\frac{\tilde{n^{-}_{e}}}{\pi m}},
\end{equation}
where $\tilde{n^{-}_{e}}$ is the electron number density at the
moment of emission. It can be slightly less than the initial
number density in the tail since during the magnetic field reconnection the
volume of emitting current sheet can widen due to some
instabilities (Zenitani \& Hoshino (2007)). We shall
adopt $\tilde{n^{-}_{e}}=\alpha\tilde{n^{-}}$, where $\alpha<1$ is 
the ratio of electron number densities in the current layer and in the tail. The total
energy of generated radio emission $E$ can be estimated as the
kinetic energy of positrons enhanced by factor $(2\Gamma )^2$. One
power of factor $(2\Gamma )$ appears due to the Doppler effect,
and the second power is due to transformation of the
kinetic energy of positrons to the frame co-moving with electronic plasma. Below
we shall not distinguish between Lorentz-factors of
accelerated electrons and positrons. Then
\begin{equation}
E=\zeta\Gamma N^+mc^2(2\Gamma)^2 \approx 4\zeta \Gamma^3 \tilde{n^+} \pi r^{*2}hmc^2,
\label{totenergy}
\end{equation}
where the factor $\zeta<1$ takes into account the efficiency of
the kinetic energy conversion into electromagnetic radiation, the
efficiency of particle acceleration etc. The $N^+$ is the total number of
positrons, $h$ is the length of the tail. The observed energy flux
is then 
\begin{equation}
S=\frac{E}{\tau \Delta_{\nu}\pi (0.5\theta r)^2}=\frac{16\zeta\Gamma^5 \tilde{n^{+}}r^{*2}hmc^2}{\tau \Delta_{\nu} r^2},
\label{obsflux}
\end{equation}
where $\theta=1/\Gamma$ is the relativistic aberration angle, $\Delta_{\nu}$ is the frequency band of the receiver, $r$ is the
distance to the source.

Taking the millisecond radio burst parameters
 $\nu\approx1.4$~GHz, $S \approx 30$~Jy, $r \approx 500$~Mpc, $\Delta_{\nu} \approx 300$ MHz (Lorimer et al. 2007) and assuming
$\tilde{n^+}=\beta\tilde{n^-}, \beta<1, \tau\approx r^*/c$ from Eq. (11)-(\ref{obsflux})
we find
\begin{multline}
\Gamma \approx 2000\left(\frac{S}{30\mbox{Jy}}\right)^{1/3}\left(\frac{h}{10^9\mbox{cm}}\right)^{-1/3}\left(\frac{r}{500\mbox{Mpc}}\right)^{2/3}\cdot\\\left(\frac{r^*}{10^8\mbox{cm}}\right)^{-1/3}\left(\frac{\nu}{1.4\mbox{GHz}}\right)^{-2/3}\left(\frac{\alpha}{0.1}\right)^{1/3}\left(\frac{\beta}{0.1}\right)^{-1/3}\left(\frac{\zeta}{0.1}\right)^{-1/3},
\label{gamma}
\end{multline}
\begin{multline}
\tilde{n^{+}} \approx 3\cdot10^3\left(\frac{\nu}{1.4\mbox{GHz}}\right)^{10/3}
\left(\frac{S}{30\mbox{Jy}}\right)^{-2/3}\left(\frac{h}{10^9\mbox{cm}}\right)^{2/3}\cdot\\\left(\frac{r}{500\mbox{Mpc}}\right)^{-4/3}\left(\frac{r^*}{10^8\mbox{cm}}\right)^{2/3}\left(\frac{\alpha}{0.1}\right)^{-5/3}\left(\frac{\beta}{0.1}\right)^{5/3}\left(\frac{\zeta}{0.1}\right)^{2/3}\mbox{cm}^{-3},\label{dens}
\end{multline}
\begin{multline}
\tilde{n^{-}} \approx 3\cdot10^4\left(\frac{\nu}{1.4\mbox{GHz}}\right)^{10/3}
\left(\frac{S}{30\mbox{Jy}}\right)^{-2/3}\left(\frac{h}{10^9\mbox{cm}}\right)^{2/3}\cdot\\\left(\frac{r}{500\mbox{Mpc}}\right)^{-4/3}\left(\frac{r^*}{10^8\mbox{cm}}\right)^{2/3}\left(\frac{\alpha}{0.1}\right)^{-5/3}\left(\frac{\beta}{0.1}\right)^{2/3}\left(\frac{\zeta}{0.1}\right)^{2/3}\mbox{cm}^{-3},
\end{multline}
\begin{multline}
E \approx 5 \cdot 10^{32}\left(\frac{S}{30\mbox{Jy}}\right)^{1/3}\left(\frac{h}{10^9\mbox{cm}}\right)^{2/3}\left(\frac{r}{500\mbox{Mpc}}\right)^{2/3}\cdot\\\left(\frac{r^*}{10^8\mbox{cm}}\right)^{5/3}\left(\frac{\alpha}{0.1}\right)^{-2/3}\left(\frac{\beta}{0.1}\right)^{2/3}\left(\frac{\zeta}{0.1}\right)^{5/3}\mbox{erg} \approx 10^{-4}\epsilon_B.
\end{multline}
The obtained value of the Lorentz-factor does not contradict to
constraints imposed by the induced scattering in the source plasma
(Lyubarsky 2008b). The positronic plasma number density and
the total energy of radio emission accord with Eqs. (7) and (6),
respectively. 

The spectrum of the observed ms radio burst is $dS/d\nu \sim \nu^{-4\pm1}$. As we mention above, the generation mechanism of the burst is analogous to that of pulsar radio emission. Consequently, it is natural that observed spectrum is similar to pulsar radio emission 
(at the high frequency range, particularly the Crab pulsar has very similar spectrum). Many physical factors can affect the spectral shape: the wide spectrum of plasma waves
($\Delta\tilde{\omega} \sim \tilde{\omega}$), the wide energy spectrum of accelerated particles, plasma inhomogeneities, plasma refraction
near the source, extinction in the parent
galaxy, etc. The observed spectrum can be partially explained by assuming
the power-law energy distribution of accelerated particles
$\tilde{dn^-}/d\gamma \sim \gamma^{-p}$. The contribution from electrons with Lorentz-factors $[\gamma, \gamma+d\gamma]$ into
the observed emission in linear approximation can be written as
\begin{equation}
dS \sim \gamma\tilde{dn^-} \sim \gamma^{1-p}d\gamma.
\label{dS}
\end{equation}
The additional Lorentz-factor is due to the Doppler effect. Then
taking into account (11) and (18) we obtain
\[
\frac{dS}{d\nu}=\frac{dS}{d\gamma}/\frac{d\nu}{d\gamma} \sim \gamma^{1-p} \sim \nu^{1-p}
\]
This spectrum is in agreement with observations provided that
$p=4-6$,
which generally does not contradict to the energy spectrum of accelerated
non-thermal particles found in numerical simulations of
relativistic magnetic reconnection (Zenitani \& Hoshino (2007)).

There could be a problem with
the generated radio burst passing through the shock.
However, the shock can be inhomogeneous
and there can be regions with significantly
lower densities (holes) transparent for radio emission.
The upper limit to the allowed density
$\rho'$ can be estimated from the condition that the plasma
frequency of the shock matter (fully ionized)
be less than the frequency of radio waves:
\[
\rho'
 \lesssim m_{p}n_{e}'
  = \frac{\pi m_pm_e\nu^2}{e^2}
   \approx 10^{-14} \mbox{ g/cm}^3
    \ll \rho,
\]
where the $n_{e}'$ is the electron number density
corresponding to the
maximum allowed plasma frequency. Indeed,
observations of SN1987a suggested strong
inhomogeneities in the expanding envelope
allowing the early appearance of
X-ray emission from the SN remnant (Grebenev \& Syunyaev 1987).
The radio emission
can also be seen through inhomogeneities, but the observed
duration and flux should be modified by effects of
passing through such an envelope.

\section{Hard radiation generation}

Let us roughly estimate properties of the possible hard synchrotron
emission which can accompany the magnetic reconnection in the magnetospheric tail. The characteristic energy of synchrophotons
is
\[
\epsilon_{\gamma} = \hbar \omega_{\gamma} \sim \hbar \frac{eB_{t}}{mc} \Gamma^2 \approx 40\left(\frac{B_{t}}{10^6\mbox{G}}\right)\left(\frac{\Gamma}{2000}\right)^2\mbox{ keV}\,,
\]
which falls into the soft gamma-ray energy range. The upper limits
to the total energy of this radiation $E_{\gamma}$ and the flux
$S_{\gamma}$ can be estimated from Eqs. (3), (14), (16) with account
for the synchrotron intensity from one particle
$I=(2/3)(e^4B_{t}^2)/(m^2c^3)\Gamma^2$:
\begin{multline}
E_{\gamma}=IN^-=\frac{2e^4B_{t}^2\Gamma^2}{3m^2c^3}\tilde{n^-}\pi r^{*2}h \approx 4\cdot10^{30}\left(\frac{B_{t}}{10^6\mbox{G}}\right)^2\left(\frac{\Gamma}{2000}\right)^2\cdot\\\left(\frac{\tilde{n^-}}{3\cdot10^4\mbox{cm}^{-3}}\right)\left(\frac{r^*}{10^8\mbox{cm}}\right)^2\left(\frac{h}{10^9\mbox{cm}}\right)\left(\frac{\tau}{3\mbox{ms}}\right)\mbox{erg},
\label{e_gamma}
\end{multline}
\begin{multline}
S_{\gamma}=\frac{E_{\gamma}}{\pi (0.5\theta r)^2} \approx 10^{-14}\left(\frac{B_{t}}{10^6\mbox{G}}\right)^2\left(\frac{\Gamma}{2000}\right)^4\cdot\\\left(\frac{\tilde{n^-}}{3\cdot10^4\mbox{cm}^{-3}}\right)\left(\frac{r^*}{10^8\mbox{cm}}\right)^2\left(\frac{h}{10^9\mbox{cm}}\right)\left(\frac{r}{500\mbox{Mpc}}\right)^{-2}\frac{\mbox{erg}}{\mbox{s}\cdot\mbox{cm}^2}.
\label{s_gamma}
\end{multline}
Apparently, the total energy of hard non-thermal radiation is
similar to that of radio emission (17), but the flux falls short
of the sensitivity of modern detectors. So no accompanying
"gamma-ray burst" is expected, which agrees with the absence of
any gamma-ray event at the time of the observed radio burst
(Lorimer et al. 2007). Clearly, should this happen in our Galaxy or in
the Local Group, both radio and gamma-ray bursts can be observed,
probably associated with an emerging supernova.

\section{Discussion}

We have shown that a bright millisecond burst of non-thermal radio
emission with properties similar to those observed by Lorimer et al.
(2007) can be explained by the supernova shock
interaction with NS magnetosphere in a massive binary system.
The shock passing through the magnetosphere leads to the formation
of a long magnetospheric tail (Istomin \& Komberg 2002). We conjecture that
the magnetic energy stored in the tail can be transformed into the
kinetic energy of non-thermal particles due to magnetic
reconnection. If there is some fraction of positrons
(which seems reasonable if NS before the SN explosion
was at the pulsar stage),
plasma instabilities can transform
the kinetic energy of the less dense positronic beam into the
plasma oscillations in the electron beam which can then be non-linearly converted into
transversal electromagnetic waves, like in pulsar magnetospheres
(Lyubarsky 1992), thereby producing a short pulse of non-thermal
radio emission. The magnetic fields of young
NSs can vary in a wide range from
 $10^{11}$ to  $10^{15}$~G, leading to a broad range of 
magnetospheric sizes, densities of pair plasma filling the
magnetosphere, etc. The situation is even more diverse in massive
binary systems with strong stellar winds from presupernova stars.
In addition, the density and velocity of SN shocks can also vary
significantly. So all basic parameters of our model ($r^*, \tilde{n^+},
\tau$, $\Gamma$ and $\tilde{n^-}$) can be very different, and so
should be the properties (frequency, flux, duration) of possible
radio transients accompanying the magnetosphere topology
restructuring during reconnection.

Magnetic reconnection can also occur in the region with
toroidal magnetic field beyond the light cylinder of a pulsar. In
this case the current sheet is located in the rotational equatorial plane of
the pulsar. The magnetic field strength at the light cylinder
$B_l$ can be significant:
    \[
    B_l \approx B\left(\frac{R}{R_l}\right)^3 \approx 4\cdot10^5\left(\frac{B}{10^{13}\mbox{G}}\right)\left(\frac{\Omega}{100\mbox{rad/s}}\right)^3 \mbox{G.}
\]
Hence there is a substantial magnetic energy storage $\epsilon_B
\sim (B_l^2/8\pi)R_l^3 \sim 10^{35}$ erg outside the magnetosphere (taking into account that toroidal magnetic field
decreases slowly as $1/r$). A stationary reconnection of the toroidal
field was suggested by Lyubarsky (1996) to
explain hard emission of pulsars. In our scenario,
the passing shock compresses the NS magnetosphere and can 
initiate a non-stationary large-scale
reconnection of the toroidal magnetic field beyond
the light cylinder. Estimates similar to those given in
Section 3 can explain the generation of short radio burst in
two-dimensional current sheet approximation. Its properties
can be similar to those observed by Lorimer et al. (2007)
assuming model parameters
similar to those given by Eqs. (14)-(17) (provided that positrons, which excite plasma
waves, are collected from a large
volume $\sim R_l^3$). Such parameters do
not contradict all existing constraints, so this approach can be
considered as an alternative model for the millisecond radio burst.
In this case it is not needed to explain
the transparency of the shock to the emergent radio emission
since the radio pulse can be produced before the SN
shock has passed the magnetosphere and screened the source.

The search for other short radio transients is under way (Lorimer et al. 2007), and larger statistics
can be used in future to test the proposed model.

\section*{Acknowledgements}
The authors thank
Profs. B.V. Somov and  V.V. Kocharovsky for useful
discussions. A valuable feedback
from Dr. Yu. Lyubarsky is also acknowledged. The work is
partially supported by the RFBR grant 07-02-00961.

\section*{References}
Goldreich P., Julian W.H., 1969, ApJ, 157, 869\\


Grebenev S.A., Syunyaev R.A., 1987, Pis'ma Astron. Zh., 13, 1024\\

Imshennik V.S., Nadyozhin D.K., 1989, Sov. Sci. Rev., Sect. E, Vol. 8, Part 1, p. 1\\

Istomin Ya.N., Komberg B.V., 2002, Astron. Rep., 46, 908\\

Lorimer D.R., Bailes M., McLaughlin M.A., Narkevic D.J., Crawford F., 2007, Science, 318, 777\\

Lyubarskii Yu.E., 1992, AA, 265, L33\\

Lyubarskii Y.E., 1996, AA, 311, 172\\

Lyubarsky Y.E., 2008a, In 40 Years of Pulsars. AIP Conf. Proc. 983, p. 29\\

Lyubarsky Yu., 2008b, arXiv:0804.2069\\

Popov S.B., Postnov K.A., 2007, arXiv:0710.2006\\

Postnov K.A., Yungelson L.R., 2006, LRR, 6\\

Usov V., 1992, Nature, 357, 472\\

Usov V., 1994, MNRAS, 267, 1035\\

Zenitani S., Hoshino M., 2007, arXiv:0708.1000

\end{document}